# Automation Of Transiting Exoplanet Detection, Identification and Habitability Assessment Using Machine Learning Approaches

Pawel Pratyush[1] · Akshata Gangrade[1]

[1] Department of Computer Science and Engineering, Maulana Azad National Institute of Technology, Bhopal, India

## Abstract

We are at a unique timeline in the history of human evolution where we may be able to discover earth-like planets around stars outside our solar system where conditions can support life or even find evidence of life on those planets. With the launch of several satellites in recent years by NASA, ESA, and other major space agencies, an ample amount of datasets are at our disposal which can be utilized to train machine learning models that can automate the arduous tasks of exoplanet detection, its identification, and habitability determination. Automating these tasks can save a considerable amount of time and minimize human errors due to manual intervention. To achieve this aim, we conducted our study in three phases - Transiting Exoplanet Detection, Identification/Triage, and Habitability Assessment. The first phase analyses the light intensity curves from stars captured by the Kepler telescope to detect the potential curves that exhibit the characteristics of an existence of a possible planetary system. For this detection, along with training conventional models, we propose a stacked GBDT model that can be trained on multiple representations of light signal (Time Domain, Frequency Domain, and Power Spectral Density) simultaneously. The next two successive phases address the automation of identification and habitability determination by leveraging several state-of-the-art machine learning models and ensemble approaches. The identification of exoplanets aims to distinguish false positive instances from the actual instances of exoplanets whereas the habitability assessment groups the exoplanet instances into different clusters based on their habitable characteristics. Additionally, in the final phase, we propose a new metric called Adequate Thermal Adequacy (ATA) score to establish a potential linear relationship between habitable and non-habitable instances. On top of classification tasks, regression analysis is also performed for the computation of Keplar Object of Intrest (KOI) and Earth Similarity Index (ESI) scores in phase 2 and phase 3 respectively. Experimental results suggest that the proposed stacked GBDT model outperformed the conventional models in detecting transiting exoplanets. Furthermore, the incorporation of ATA scores in habitability classification enhanced the performance of models.

**Keywords**  Exoplanet Detection . Habitability Assesment . Machine Learning . Bias Handling . Regression Analysis

## 1 | Introduction

One of the most profound question that human civilization has ever pondered is 'Are there other planets like earth outside our solar system which can support life?' This question has puzzled scientists and philosophers for millennia, but we are the first generation who are privileged enough to be equipped with the proper tools and technologies that have enabled us to scientifically approach and deduce the answer to this question. The approaches for detecting the planets outside of our solar system, also known as exoplanet detection have led to several ingenious discoveries in the past few decades (Bordé et al 2003). Due to increasing global warming and ozone layer depletion, the future of the earth seems alarming. On the other hand, there is a possibility of the earth being hit by asteroids at any time in the future. We currently do not possess any sophisticated technology that can prevent an asteroid attack. Thus, there is an exigent need to discover planets in space that are capable of supporting life. This has emphasized the research on detecting exoplanets in outer space and thereby finding out evidence of habitability on those planets. Moreover, the discovery of new planetary systems enlightens us about the working of the universe and how the earth, sun, and our solar system fit into the whole picture.

Exoplanet's expedition over the years has predominantly been done by a group of highly skilled individuals specializing in astrophysics. However, NASA's recent exoplanet probe has led to the launch of several satellites such as K2, TESS, Plato 2.0, etc that are capable of garnering an extensive array of data related to different celestial objects (Akeson et al. 2013). This has opened doors to individuals who are competent in devising and elucidating machine learning models.

Over the years, a lot of techniques have been put forward in search of extrasolar planets under the umbrella of the Kepler mission. Some of these classical identification techniques include radial velocity, microlensing, direct imaging, pulsar timing, eclipse timing variation, transit, and so on (Fischer et al. 2014). However, among these techniques, transit has been the one that has identified a significant number of planets compared to other techniques[1]. Therefore it has opened a whole new window for discoveries among the scientific communities around the globe.

Conventional techniques like direct imaging and radial velocity are more inclined towards the detection of large-sized planets. The transit has an edge over these techniques due to the fact that it is scalable for the detection of smaller-sized planets (Deeg et al 2018). A transit can be defined as an identical event to that of a solar eclipse, which takes place when an exoplanet passes between the star it orbits and the observing satellite (Fig. 1). Light intensity curves emitted from stars can be used in order to study these transiting events. As depicted in Figure 2, when an exoplanet passes in front of the star, there is a decrease in the intensity of light (dip) when observed by a satellite, indicating the possibility of the occurrence of a transit event.

However, in practice, detection using transit technique has heavily relied upon manual inspections of a large corpus of data which is inexorably prone to human error and thus false positives are most likely to eventuate (Osborn et al 2016). The removal of false positives is therefore critical which is yet another point of interest of our research. To account for this, the obtained exoplanets from transit methods need to be cross verified upon further analysis. This is done by a vetting process called triage in which threshold crossing events (TCEs), which is a series of transit-like features in the light curve of a particular star, are examined. The Kepler Objects of Interest (KOI) from suspicious curves of TCEs that contain some instrumental noise are thereby separated. A KOI is a TCE that does not exhibit any blatant proof that the event was prompted by a non-transiting anomaly and thus it is passed on for further analysis. This analysis can be performed by leveraging KOI tables which contain cumulative information of individual KOI activities that provide the most reliable dispositions and stellar and planetary information with the help of which it classifies each transiting planet as a candidate or false positive (Morton et al. 2016). Once the detected exoplanets are confirmed as a candidate planet, the ultimate goal is to find out the answer to the question "Whether these planets can sustain the life of living organisms and support the advancement of human civilization?". With the launch of the ESA Gaia satellite observatory in 2013 and NASA's Kepler mission in 2009, scientists have been able to collect data related to the atmospheric and physical properties of celestial objects. This data can be promising in determining habitability with the aid of computational and Artificial Intelligence methods.

---

[1] https://exoplanetarchive.ipac.caltech.edu/docs/counts_detail.html

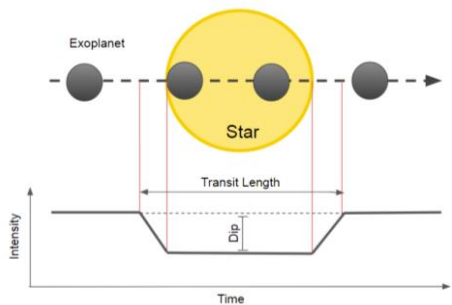

**Fig**. **1** Transit Event

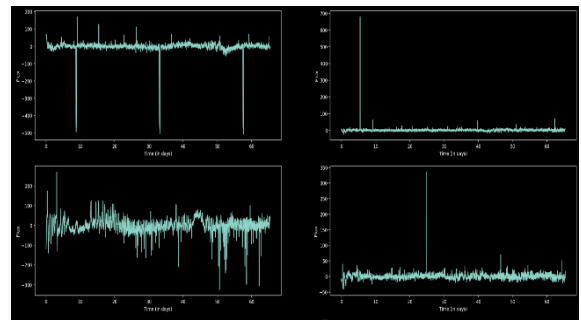

**Fig**. **2** Light Curve Samples of Exoplanet (left) and Non-Exoplanet (right)

The paper is organized as follows: Section 2 contains the previous work done in the field of exoplanet detection, identification and habitability assesment. Section 3 explains the methodology of the research. Section 4 contains the description of datasets used to train and test the models proposed in this work. Section 5 explains the preprocessing techniques that are applied on the datasets. Section 6 explains the first phase i.e Transitory signal detection. Section 7 explains the second phase i.e Transitory signal identification. Section 8 explains the third phase i.e Habitability Assesment. Evaluation Procedure for performance analysis is discussed in Section 9. Results of the employed models and their comparison is discussed in Section 10. Conclusions drawn from the research are included in Section 11.

## 2 | Related Work

Various methods have been proposed by researchers for the detection of exoplanets using AI-driven methods. In a broader sense, these methods can be looked at as a pipeline of three steps: 1) acquisition of light signal 2) preprocessing 3) detection and identification of transit signal. The acquisition of signals is done by using sophisticated tools like the Kepler satellite or through simulation methods (Kumar et al. 2020). A significant amount of research has been done which addresses how machine learning algorithms can be utilized to filter out noises induced in the signal that may occur due to instrumental error or other external factors. For instance, Grziwa et al. (2016) proposed a hybrid model by integrating the trend filtering approach with least square harmonic filter. This proposed model focused on enhancing the identification rates by permitting to detect signals with larger span and dull transits. Author Smith et al. (2012) proposed a Bayesian approach incorporating Bayesian Maximum a Posteriori (MAP) which targeted eliminating device-generated error while safeguarding the source of research interest such as stellar heterogeneity. MAP model is employed to trust the prior knowledge, based on the amount of stellar heterogeneity observed. Tamuz et al. (2005) leveraged Principal Component Analysis (PCA) for reducing the dimensions of the signal features by finding out the correlation between them. An iterative process was then used to select a model that gave the best performance in removing the noise from the light signals.

Once the removal of noise from light signals is done, the next step is the detection and identification of transiting planets from these signals. In this regard, Pearson et al. (2018) used CNN to detect exoplanets from light signals where the network is trained with the synthetic photometric signals by adding an array of various parameters like depth of transit and orbital cycle. whereas, Author Chintarungruangchai et al. (2019) used a multi-dimensional CNN that presents a novel way of signal folding for the detection of exoplanets. Pearson et al. (2018) trained a neural network after converting the light signal into the frequency domain which enabled the neural net to learn from the important features of signals whilst disregarding the least contributing features including noise.

For identification of detected exoplanet, Coughlin et al. (2016) proposed a tree-based model termed as Robovetter which performs several tests by employing decision trees. This model makes decisions on the basis of threshold crossing event's span, multiple event statistics, and maximum single event statistic values along with the inclusion of the planet's radius. Locality Preserving Projections (LPP) is a dimensionality reduction method used by Thompson et al. (2015) along with k-nearest neighbors to identify exoplanet signals. This dimensionality reduction technique was more effective to outliers than the other techniques such as PCA. The Euclidean distance was used as a similarity measure to cluster transit-like signals. A CNN called Astronet was proposed by Shallue and Vanderburg (2018) to classify potential planet signals. It was compared with Artificial Neural Network (ANN) with zero hidden layers, a Fully Connected Neural Network , and a 1-D CNN. The obtained results show that CNN performs best among all the models and thus, CNN was used to identify Kepler TCE.

The vast data collected by Gaia and Kepler has recently attracted a lot of interest in determining habitability on potential exoplanets using various Machine Learning approaches. Basak et al. (2018) used PHL Catalog data to characterize the habitability of exoplanets into different classes. This classification is performed in two steps - first using a synthetic oversampling technique called Parzen Window Estimation and then employing machine learning classifiers such as Probabilistic (Gaussian Naive Bayes), Instance-Based (KNN), Hard-boundary (SVM), and Tree-based (CART and Random Forest) on the synthetic samples. The results were then interpreted for each classifier. A splitting criterion was also introduced for tree-based algorithms called Elastic Gini. It was found that the tree-based classifiers performed the best followed by SVM. Bora, K et al. (2016) used analytic modeling to propose a new habitability metric called Cobb–Douglas Habitability Score (CDHS). CDHS calculates habitability score by utilizing measured and estimated input planetary features. The calculated CDHS scores are then given to the KNN model to classify exoplanets to suitable habitable classes and thus creating different group of habitability.

## 3 | Research Methodology

A high-level view of the proposed architecture can be seen from Figure 3. This architecture is constructed by dividing our study into three phases where each phase serves a goal to automate the concerned task. These phases are arranged in a pipelined fashion and are described below:

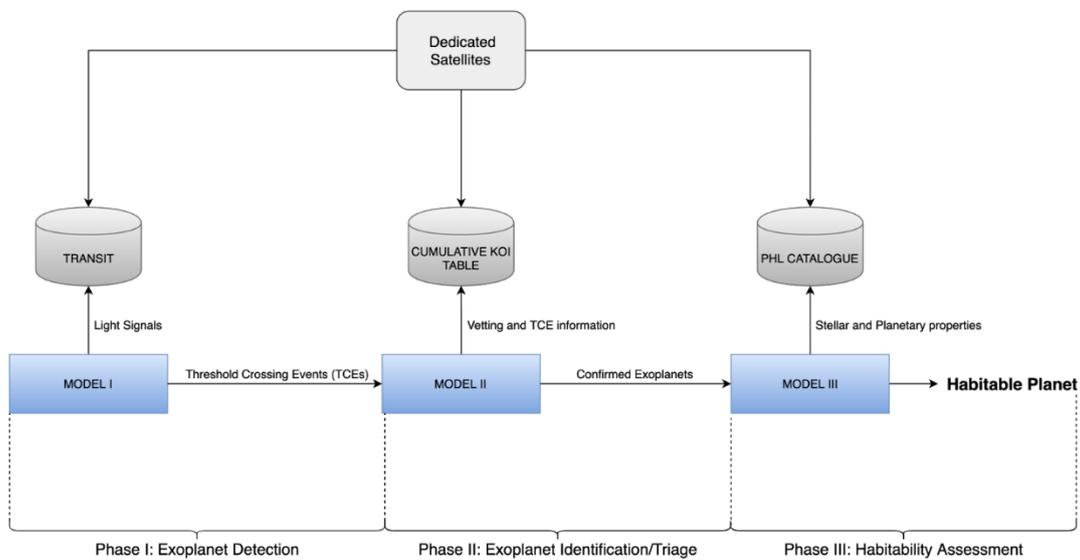

**Fig**. **3** High-level view of the wor

*Phase I:* This phase is called the Transitory Exoplanet Detection phase. Here, the light signals (or light curves) are used to train classifiers to detect the significant dip occurred in the light intensity of stars. The detection of dip in the light curve of a star may characterize a transit phase which in turn gives us the possibility of existence of a planetary system around that star. The light signals data cumulated by satellite may be vulnerable to instrumental noise and other factors like unscaled attributes and subtrends causing distortion. Hence, detection is done after appropriate preprocessing of the signals. The preprocessing steps undertaken in this phase are described in Section 5(a). These signals are originally in the time domain and are converted into other domains such as frequency domain and spectral power density where they may possess different characteristics. Next, the converted signals in each domain are subjected to processes of dimensionality reduction to eliminate the curse of dimensionality followed by bias handling to deal with the imbalance in the dataset which are dealt in Subsections 6.2 and 6.3 respectively. The model performance of various classification models is then compared to observe if the machine learning models are capable of efficiently automating the process of detecting an exoplanet system around a star using transit signals. The classification phase is dealt in Subsection 6.4

*Phase II*: This phase is called as Exoplanet Identification (or Triage) phase. Here, the possible Threshold Crossing Events (TCEs) are analyzed for suspicious light signals that were identified as Exoplanets in the detection phase. To account for this, the cumulative KOI table contains necessary vetting and TCE information to investigate events related to various KOI objects. This information is utilized to classify whether the suspected signals are of actual exoplanets or false positives due to instrumental noise. The preprocessing steps performed on KOI tables are described in Section 5(c). The analysis and comparison of several classification models to automate this phase is discussed in Subsection 6.4.

*Phase III*: This phase is called as Habitability Assessment phase. This phase classifies habitable characteristics of exoplanets into granular clusters of Conservatively habitable, Optimistically habitable, and Non-habitable. The preprocessing steps applied to the given dataset are discussed in section 5(b). A new metric called Absolute Thermal Adequacy (ATA) score is added in the dataset that can provide useful information for learning the classifiers for habitability classification which is described in Subsection 8.2. The classifiers are trained after bias handling followed by feature selection. These two steps are dealt in Subsections 6.2 and 6.1 respectively.

In addition, regression analysis of fundamental features in identifying exoplanet (Phase II) and determining habitability (Phase III) is also performed which are discussed in Sections 7 and 8 respectively.

## 4 | Dataset Description

We have utilized three datasets for our study, each of which is used in the relevent phase of our architecture. These datasets are described as follows:

***Transit Data***: This data is hosted by CalTech in association with NASA which can be obtained from exoplanet archive[2]. The dataset that we extracted consists of light curves of more than 5500 stars which have been periodically collected after every 29.5 minutes by the K2 mission over the span of around 2000 hours. Each row in the dataset denotes the vector of light intensities of a particular star. Since the phenomenon of finding an exoplanet is very rare, this data is highly imbalanced and it consists of only less than 1% labeled instances of candidate exoplanets (Fig. 4a).

***KOI data***: The data comprising the KOI attributes - Cumulative KOI table is also hosted by the joint initiative of Caltech and NASA[2]. The dataset contains stellar object identifiers, mission disposition columns, exoplanet archive attributes, threshold-crossing events, condensed transit features, stellar variables, and KOI vetting information based on photometric pixels of stars. There are 9564 rows each representing information about a Kepler Object of Interest. The target label corresponding to each row can be classified either as false positive or candidate exoplanet.

***Habitability data***: This data is hosted by Planetary Habitability Laboratory which is a joint venture of the University of Puerto Rico with the collaboration of NASA and SETI[3]. The data consist of aggregate level information about the planetary properties (planet's radius, mass, eccentricity, equilibrium temperature, earth similarity index, etc) and stellar properties (star's age, magnitude, effective temperature, metallicity, etc) having 4048 rows, each representing an exoplanet with its corresponding planetary and stellar properties. A habitability assessment column called 'P_HABITABLE' consisting of three discrete values - Not Habitable, Conservatively Habitable, Optimistically Habitable, is treated as a target variable. Similar to transit data, this data also presents a high imbalance nature since non-habitable instances constitute more than 99% of data (Fig. 4b).

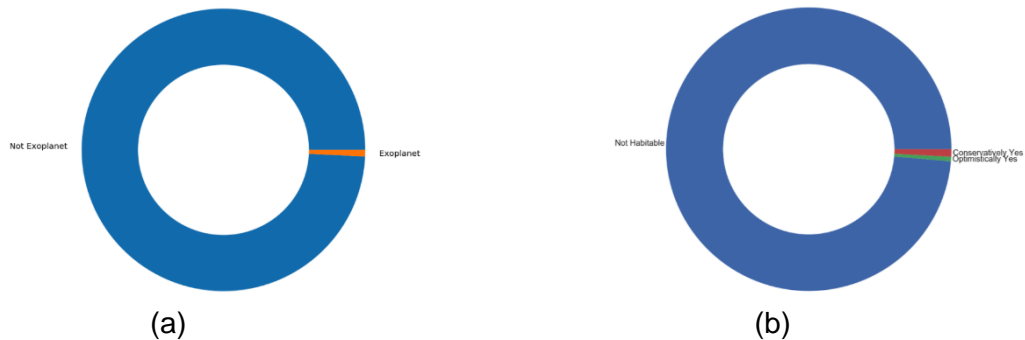

(a)          (b)

**Fig.** 4 Class distribution **a:** Transit Data **b**: Habitability Data

## 5 | Dataset Preprocessing

The data used in this paper possess many challenges for training robust machine learning algorithms. A robust machine learning algorithm can only be trained if the class distribution is uniform and the data is structurally dense (Wei et al. 2013). Improper class distribution leads to bias towards the class having a higher share of samples. Similarly, sparsity in data affects the learning of models and leads to poor performance on test data. Other challenges that hinder the performance of models are unscaled features, noise, and uncorrelated features. Bias handling of imbalance data is discussed in Subsection 6.3. The other preprocessing steps are described in the following subsections:

*(a) Transit data*: Light curves of stars was initially plotted from the transit data for the time span of roughly 65 days. It can be observed from Figure 5 that the scale is not uniform for different light curves which can hinder the convergence of optimization-based classifiers. Hence, the signals were first scaled to Z-score for coherence. Moreover, some stars showed linear trends during their time span which might not contribute considerably in training classification models. So, each light curve was divided into fifty equal segments and then detrending was performed piecewise on each one of them. Figure 5 shows the result of scaling and detrending of signals. Since the signals can be affected by instrumental noise, the next step undertaken was noise removal. An appropriate filter needs to be selected that can have a pronounced effect on noise removal with minimum loss in details. For this, we have adopted three popular filters - Uniform filter, Gaussian filter, and Stavisky Golay Filter. Figure 6 shows the effect of applying these filters on a light curve. We can see that the Gaussian filter has produced the sharpest signal as compared to the other two filters. Nonetheless, it might also have caused a loss of information. As a consequence, we have implemented these filters independently and the one delivering the best results was selected as the final filter.

---
[2]https://exoplanetarchive.ipac.caltech.edu/cgi-bin/TblView/nphtblView?app=ExoTbls&config=cumulative
[3]http://phl.upr.edu/projects/habitable-exoplanets-catalog/data/database

*(b) PHL data*: This data suffers from significant sparsity which can be confirmed from the data pattern shown in Figure 7(a). To resolve this, we first calculated the percentage of missing values in each column and decided a threshold as 30. We dropped those columns whose percentage of missing values exceeded the threshold (Fig. 7(b)). Some of the features were perfectly correlated pairwise as illustrated in Figure 7(c). This may suggest that these features are duplicated and hence we kept one column from these pairs and dropped the other one (Fig. 7(d)). The remaining missing values were then imputed for both categorical features and numerical features. Only three categorical features contained missing values which were calculated using the mode strategy. Numericals features were obtained using Multivariate Imputation using Chained Equations (MICE). The data was finally scaled into a standard normal distribution before passing it to the classification phase.

*(c) KOI data:* This data shares the similar structural characteristics as PHL data. Athough the sparsity in data is significantly less compared to the PHL data, the preprocessing steps are similar in both the datasets.

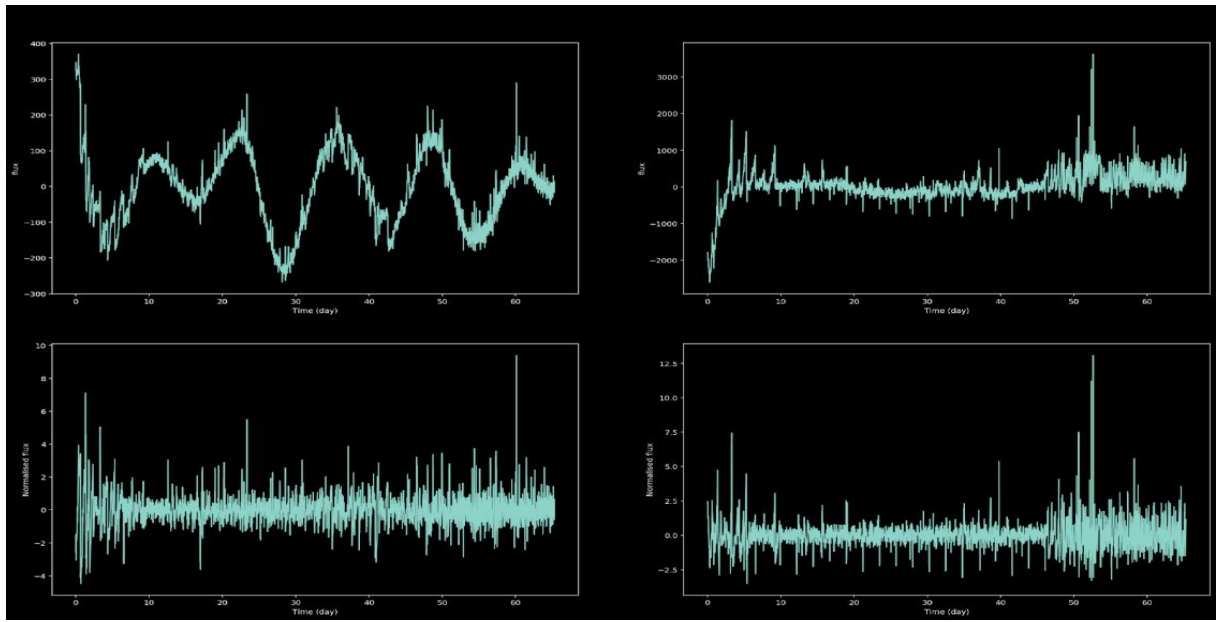

**Fig**. 5 The upper row represents original light signals and the lower row represents light signals obtained after scaling and detrending

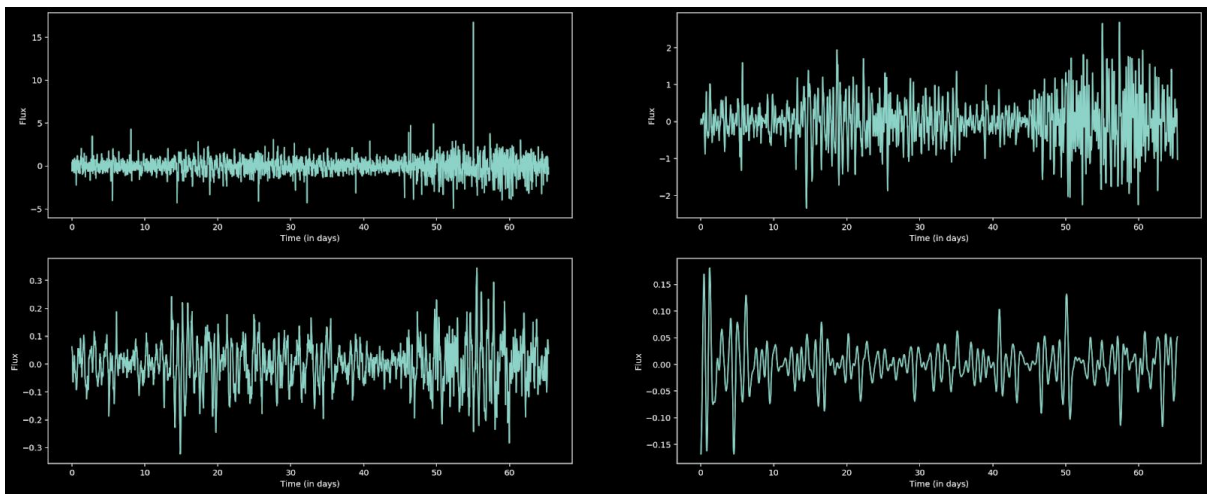

**Fig**. 6 Original Signal (upper left), after using Uniform Filter (upper right), Stavisky Golay (lower left) and Gaussian (lower right)

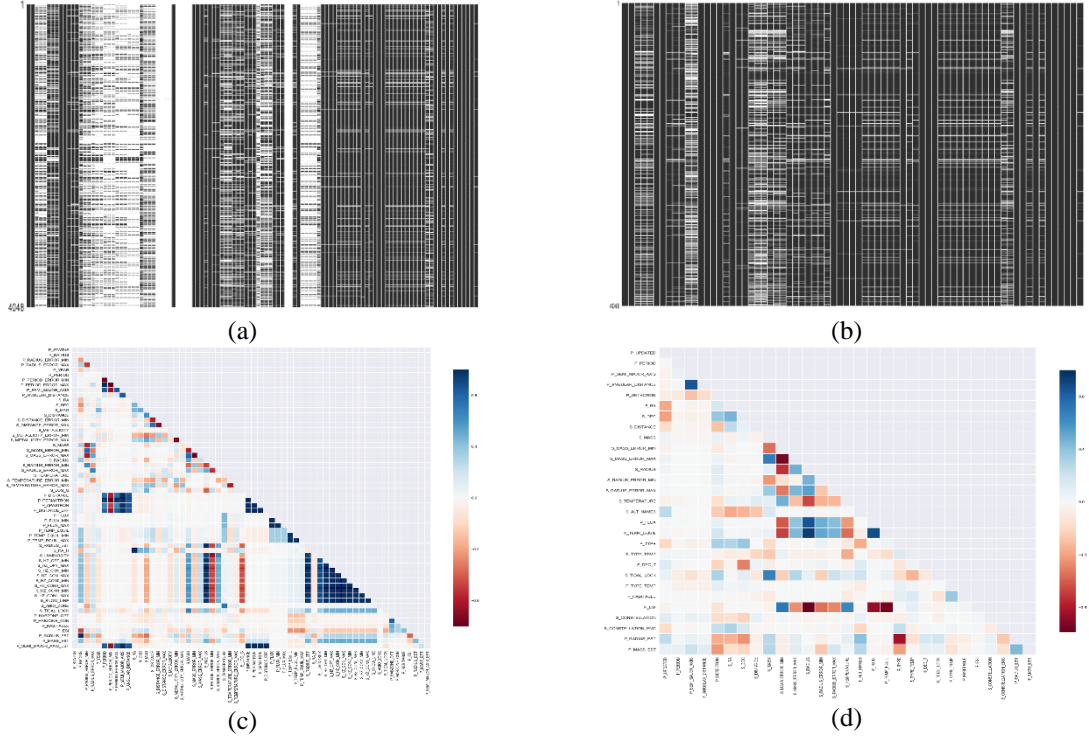

**Fig**. 7 missing value pattern in dataset **a**: before imputation **b**: after imputation. Heatmaps **c:** before dropping features **d**: after dropping perfectly correlated feature

## 6 | Transitory Signal Detection (Phase I)

The flowchart below (Fig. 8) represents the sequence of processes that are involved in transitory signal detection. In a broader view, mainly four processes are executed - Feature Extraction, Feature Reduction, Bias Handling, and Classification.

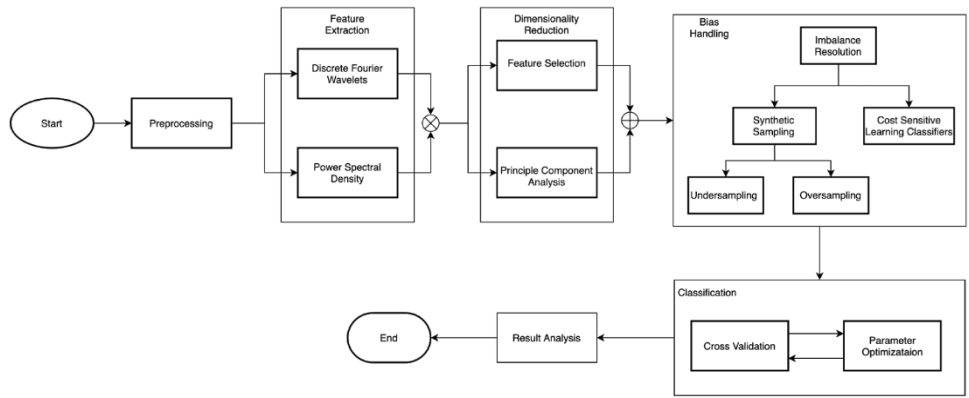

**Fig**. 8 Flowchart of exoplanet detection

**(1) Feature Extraction**
The light curves are light intensity signals represented as a function of time. Obtaining these signals in the frequency domain using Discrete Fourier Transform (DFT) provides a simpler representation of the signal than the time domain representation (Fig. 9). One of the main advantages of obtaining signals in the frequency domain is that it can improve the signal-to-noise ratio (SNR) (Zhang et al. 2016).

Let $x(n)$ be the input sequence of light intensity signal representing the analogous continuous time signal $x(t)$, the Discrete Time Fourier Transform (DTFT) of this sequence is represented by $X(e^{j\omega})$ which is a periodic function in $\omega$ with a period of $2\pi$.

$$X(e^{j\omega}) = \sum_{n=-\infty}^{+\infty} x(n)e^{-jn\omega} \quad (1)$$

Here, the aim is to get the set of sinusoids which can be summed up together to get $x(n)$. For this, we take N samples of signals from each period of $X(e^{j\omega})$. The frequency of the set of sinusoids will be of the form $2\pi N * k$, where $k = 0,1, \ldots N-1$. Now, by using complex exponentials, $x(n)$ can be written as:

$$x(n) = \sum_{k=0}^{N-1} X'(k) e^{-j\frac{2\pi}{N}kn} \qquad n = 0,1,\ldots, L-1 \tag{2}$$

On sampling this DTFT at evenly spaced frequency values, we get the DFT of light signals. In order to calculate this DFT of a signal having a period of N, we first need to convert it into periodic signal $p(n)$ from $x(n)$. Considering that period of signal $(p(n))$ equals that of the input signal $(x(n))$ for $n = 0,1, \ldots N-1$, DTF of this periodic light signal is:

$$\alpha_k = \frac{1}{N} \sum_{n=0}^{N-1} x(n) e^{-j\frac{2\pi}{N}kn} \tag{3}$$

On multiplying the coefficients of the Equation 3 by period N, the DFT coefficients, $X(k)$, can be derived as:

$$X(k) = \sum_{n=0}^{N-1} x(n) e^{-j\frac{2\pi}{N}kn} \tag{4}$$

It is important to note that although the discrete-time fourier series of a light signal is periodic, the DFT coefficients, $X(k)$, are a finite-duration sequence defined for $0 \leq k \leq N$.

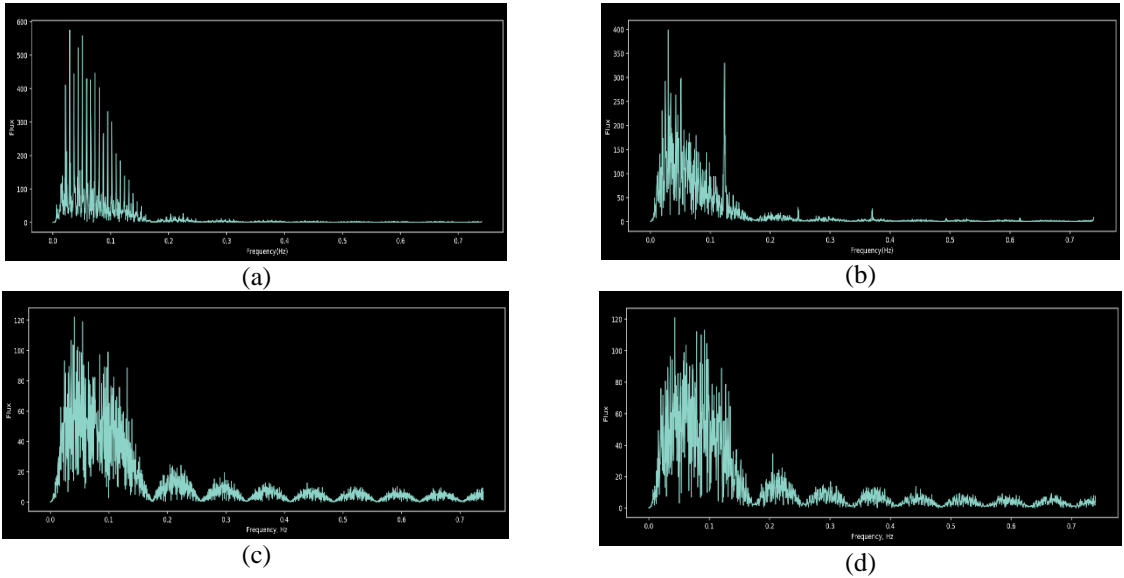

Fig. 9 Signals in frequency domain **a-b**: Non-exoplanet **c-d**: Exoplanet

The other way of transformation is to represent the spectral energy distribution of the light signal as a function of the signals' frequency. This representation is called Power Spectral Density (PSD). PSD function helps us in identifying oscillatory signals in time series data and thereby obtaining amplitudes of these signals. PSD can be used to learn relevant patterns in light signals by identifying strong variation ranges of frequency (Fig. 10).

The power spectral density of light signals can be obtained by using DFT of these signals. To do so, we consider a light signal $x(t)$ in the time domain which is sampled at every interval of t seconds. The total time length of the sampling window containing $N$ samples is $N\Delta t$. The PSD in terms of DFT is given as:

$$S(f) = \frac{1}{N} |\sum_{n=1}^{N} x(\Delta t) e^{-i2\pi f n \Delta t}|^2 \tag{5}$$

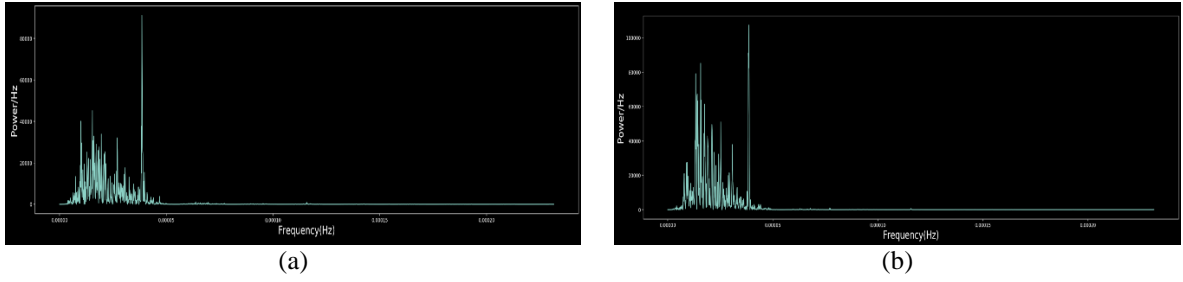

Fig. 10 Power spectral density **a**: Exoplanet **b**: Non-exoplanet

**(2) Dimensionality Reduction**

Each of the light curves consists of thousands of features. If we consider all of the features, it can lead to expensive computation as well as overfitting and we may end up getting misleading results on the test dataset. To work on this, we adopted two different techniques for reducing the number of features in the dataset. These techniques are discussed as follows:

**(a) Feature Selection**: This is done by ranking the features on the basis of Kendall Tau correlation coefficient with respect to the target variable. The reason for selecting Kendall Tau as the interdependence measure is that all the features in this data are continuous and are non-linear in nature (Puka L. 2011). After performing the ranking of features, we selected the K best feature subset from it. The K value was hypertuned using GridSearch and the final K features were selected as the input to the classification phase.

Let $x$ be the vector of light intensity values of all stars at a particular interval of time and $y$ be the vector of output labels corresponding to each star, each of size n. On considering the ordered pair of values $x$ and $y$ i.e. $(x_i, y_i)$ and $(x_j, y_j)$, if the sort order of this pair of stars agrees (either $(x_i > x_j$ and $y_i > y_j)$ or $(x_i < x_j$ and $y_i < y_j)$), then the pair is concordant, otherwise the pair is discordant. The Kendall tau coefficient between x and y can now be calculated by the formula below:

$$\tau = \frac{(n_c - n_d)}{\left(\frac{n(n-1)}{2}\right)} \quad (6)$$

where, $n_c$ number of concordant pairs, $n_d$ = number of discordant pairs

**(b) Principal Component Analysis**: This is another technique from which we can extract prominent features with respect to the target variable from data and train on those extracted features only. These features are obtained by generating new uncorrelated features of transit signals that successively maximizes the variance and thus reduces the problem to compute eigenvectors/eigenvalues.

Let $X$ represent the light signal data matrix, $\bar{X}$ be the vector of the average of standardized light intensity values of all stars at each interval and $cov(X)$ be the covariance matrix of $X$ with respect to target variable which is calculated using the formula:

$$cov(X) = \frac{1}{n-1} \sum_{i=1}^{n} (X_i - \bar{x})(Y_i - \bar{y}) \quad (7)$$

where $X_i$ is the intensity values of the $i^{th}$ star at a particular interval and $Y_i$ is the corresponding values of the target variable at that interval. $\bar{x}$ is the mean value of the light intensity values of n number of stars whose value can be taken from vector $X$. $\bar{y}$ is the mean of target variable. Now, we calculate eigenvectors of the matrix $X$ whose eigenvalues are the roots of the characteristic equation:

$$\det(X - \lambda I) \quad (8)$$

Here, $I$ is an identity matrix of the same dimension as of $X$. By using eigenvectors, we are transforming the covariance matrix into a diagonal that gives us the maximum variance with minimal correlation between new components. Next, the eigenvectors are sorted by descending values of eigenvalues and top k values are selected and concatenated together to form a matrix $W$. Finally, the light signals are transformed into new-subspace $y'$ by using by the operator:

$$y' = X * W \quad (9)$$

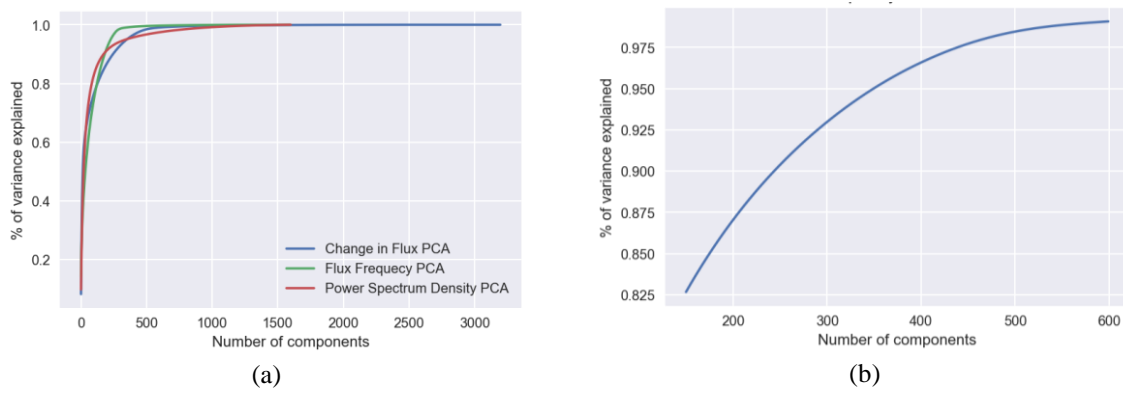

(a)                  (b)

**Fig**. 11 Percentage of variance explained corresponding to the number of components **a**: For each characteristics **b**: Frequency PCA knee

Interestingly, the explained variance raises more rapidly in the frequency domain than in the time domain (Fig 11(b)). We would barely need 250 components to have more than 95% of explained variance (Fig. 11(a)). Hence, the size of feature set is reduced from 3196 to only 250.

## 3. Bias Handling

Uneven distribution of class labels might create a bias towards majority class labels (Leevy et al. 2018). Transit data is a case of severe imbalance where the distribution of labels in the training set is uneven by a large amount. Most classification predictive models work by assuming an equal distribution of output labels. This means that a naive application of a machine learning model may focus on learning the characteristics of the abundant observations and thereby neglecting the examples from the minority class that is, in fact, of more interest and whose predictions are more valuable. We have undertaken three different strategies - oversampling, undersampling, and cost-sensitive learning to resolve dataset imbalance.

**(a) Oversampling:** Oversampling methods duplicate examples in the minority class or synthesize new examples from the examples in the minority class. We have employed Adaptive Synthetic (ADASYN) as an oversampling technique. ADASYN makes use of weighted distribution for different minority class instances on the basis of their difficulty level in learning. In this way, for those minority signal instances that are harder to learn, more synthetic data is generated than those that are easier to learn. Hence, this approach refines learning by adaptively shifting the classifier decision boundary towards the difficult instances

.**(b) Undersampling**: These techniques remove examples from the training dataset that belong to the majority class in order to balance the class distribution. It is important to note that undersampling might also remove majority class samples that carry essential information that may be quite useful for learning classification models. It is therefore a crucial decision to choose an appropriate undersampling algorithm that preserves important information. In this regard, we adopted One Sided Selection algorithm (OSS) to perform the undersampling of data.

**(c) Cost Sensitive Learning**: A cost matrix codifies the penalties of misclassifying examples. Let C be a cost matrix where C(i,j) denotes the cost of predicting class i example as class j; Thus, in this notation, C(+ve, ) is the cost of misclassifying an example of the positive class as negative and C( ,+ve) is the cost in the opposite scenario. In case of imbalanced classification, recognizing the positive examples is more vital and thus, the misclassification cost of a positive example is higher than that of a negative one (C(+ve, ) > C( ,+ve)). Section 6.4 elucidates various cost-sensitive models that we have leveraged in this paper.

## 4. Classification

The balancing of the dataset is followed by the classification of the light signals. Clearly, the scatter plots in Fig. 12 shows that, with these features as is, there exist no linear decision boundary that separates the two targets. While this observation is only based on a subset of features that we looked at, it gives a good starting point to investigate models that make use of non-linear decision boundaries. Therefore, we employed kernelized models like RBF-SVM with non-linear mapping activation function, along with tree-based classifiers which are capable of addressing non-linear relationships among features. These classifiers are discussed briefly in the following subsections:

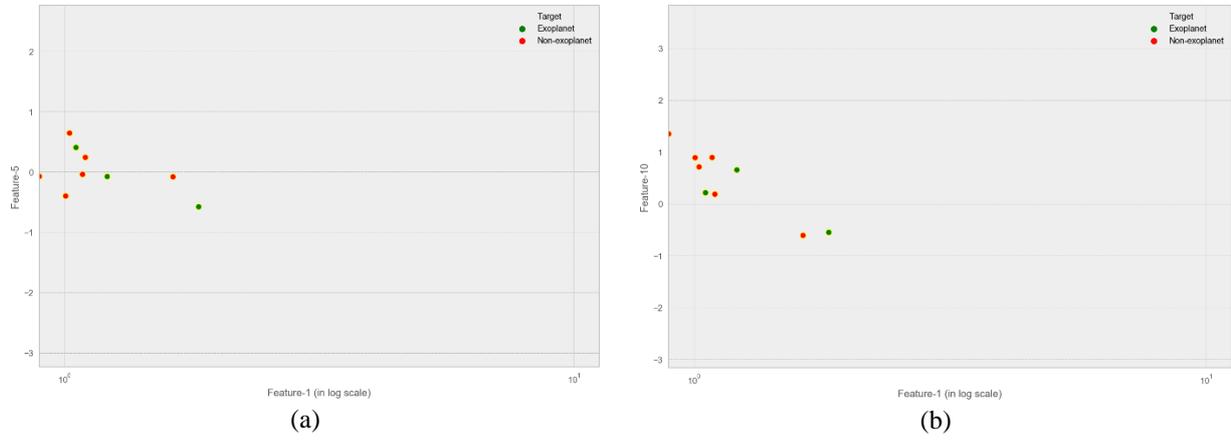

**Fig**. 12 Scatter plots  **a**: Feature 1 vs Feature 2  **b**: Feature 1 vs Feature 5

(Feature 1 was scaled to log in each of them)

**(a) SVM:** Support Vector Machine (SVM) is trained to find a hyperplane that separates data into two classes. It does so by finding out the maximum margin between two closest training examples of each class. In case of non-linearly separable data, Soft margin SVM allows some instances to violate margin constraint but with some penalty imposed on them. This problem of constraint optimization can be formulated as:

$$\min_{w,b,\xi} (\frac{1}{2}||w||^2 + C\sum_{i=1}^{m}\xi_i) \qquad (10)$$

$$s.t. \quad y_i(w^T\phi(x^i) + b) \geq 1 - \xi_i \qquad (11)$$

$$\xi_i \geq 0, \quad i = 1, \ldots, m \qquad (12)$$

Here, $w$ and $b$ are the weight and bias term vectors respectively, $\xi$ is the slack variable and $C$ is the regularization constant that controls the tradeoff between margin width and misclassifications. The dual of the above primal problem boils down to:

$$\min_{\alpha} (\frac{1}{2}\sum_{i,j=1}^{m}\alpha_i y_i y_j K(x_i, x_j)\alpha_i - \sum_{i=1}^{m}\alpha_i) \qquad (13)$$

$$s.t. \quad \sum_{i=1}^{m}\alpha_i y_i = 0 \qquad (14)$$

$$0 \leq \alpha_i \leq 1, \quad i = 1, \ldots, m \qquad (15)$$

Here, $K$ is a kernel used for mapping non-linear training data into a higher dimension using function kernel function $\phi$. In this work, we have used RBF kernel function as $\phi$ which does this transformation by finding out similarity between the data points :

$$where \quad K(x_i, x_j) = \phi(x_i)^T\phi(x_j) \qquad (16)$$

$\gamma$ defines how far the impact of a single training instance reaches.

**(b) Decision Trees:** A Decision Tree is a powerful decision support tool that is used for classification and prediction. It uses a tree-like model of decisions and their possible consequences. In this tree, each internal node denotes a test on an attribute, each branch represents an outcome of the test, and each leaf node holds a class label. It is a non-parametric method that creates a model which learns decision rules deduced from the data features and applies these rules to predict the value of the target variable.

**(c) Random Forest**: From the set of training data, a bootstrap sample is drawn randomly with replacement, and this sample is then used to build each tree of the random forest (first layer of randomness). During the construction of the tree, each node is split and the best split is decided either from all input features or from a random subset of input features (second layer of randomness). The individual tree in the forest shows high variance and tends to overfit. Therefore, incorporating these two layers of randomness abates the forest estimator's variance.

**(d) Extra Trees**: Here, instead of selecting those thresholds that are most discerning, another layer of randomness is added as thresholds are randomly chosen for every candidate feature, and out of these, the best random thresholds are picked as the splitting rule. Thus, this cost of slightly increasing the bias results in reducing the variance of the model.

**(e) Gradient Boosting**: Gradient boosting converts weak learners into stronger learners by performing the ensembling of weak prediction models. Like other boosting methods, gradient boosting trees are also built iteratively in a stage-wise manner, but it differentiates from the other methods as it performs the optimization of the arbitrary differentiable loss function by choosing a weak hypothesis that points in the direction of negative gradient.

## 5. Proposed Stacked Gradient Boosting Decision Trees (Stacked GBDT) Model

A stacked GBDT model is proposed which is capable of utilizing multiple characteristics of the light signal. The idea behind this model is to train not only on a single set of features (flux) but on a multitude set of features simultaneously each bringing crucial information to the learning algorithm.

Let $x_0$ be the feature vectors of light signal in time domain (flux), $x_1$ be the representation of flux features into its corresponding frequency domain (Discrete fourier wavelets) and $x_2$ is the feature vector obtained by transforming the flux features into spectral energy distribution (Power Spectral Density). Let $[y_0, y_1, y_2]$ be their corresponding predictions obtained after feeding each into the GBDT classifier. Mathematically, for each input characteristic feature vector, we minimize the loss function $L(y_i, F(x_i))$ with $F(*)$ being ensembled trees.

$$F^* = \min_F l(y_i, F(x_i)) \ with \ F(x) = \sum_{m=1}^{T} f_m(x) \tag{17}$$

here, $f_m$ denotes decision tree. The loss function is approximated by quadratically as:

$$\sum_{i=1}^{n} l_i(\hat{y}_i) + f_m \approx \sum_{i=1}^{n} (l_i(\hat{y}_i) + g_i f_m(x_i) + \frac{1}{2} h_i f_m(x_i)^2) \tag{18}$$

$$= \sum_{i=1}^{n} \frac{h_i}{2} ||f_m(x_i) - g_i/h_i||^2 + c \tag{19}$$

where $g_i = \partial_{\hat{y}_i} l_i(\hat{y}_i)$ is the gradient and $h_i = \partial^2_{\hat{y}_i} l_i(\hat{y}_i)$ is the second order derivative

Hence, the objective is to find $f_m(x, \theta_m)$ that minimizes the following loss function,

$$\min_{f_m} \sum_{i=1}^{N} [f_m(x_i, \theta) - g_i/h_i]^2 + R(f_m) \tag{20}$$

XGBoost suggests that computing the second order derivative gives us better performance (Chen et al. 2016).

Now, each of these predicted vectors ($y_0, y_1, y_2$) are stacked together and then fed to the Multi Layer Perceptron (MLP) model. The final predictions are eventually collected from the output of the MLP model. If $p_i$ be the predicted probability values of the MLP model and $q_i$ be the corresponding inputs from $y_0, y_1, y_2$, the model training boils down to just minimizing cross entropy loss function given by:

$$L_{CE} = -\sum_{i=1}^{N} p_i \log q_i \tag{21}$$

In this way, the model is capable of training on multi-characteristic feature sets simultaneously. The architecture of proposed stacked GBDT is shown in the Figure 13:

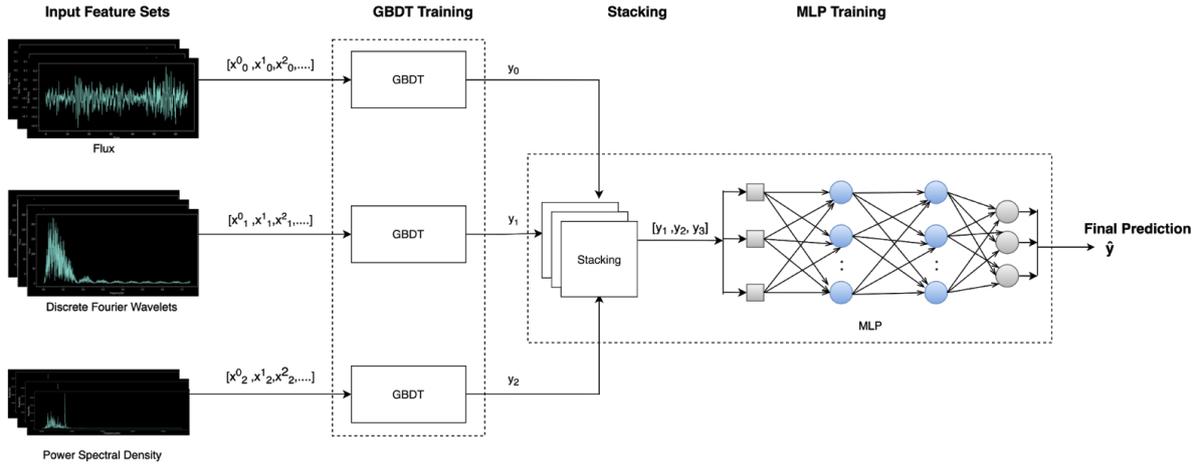

**Fig. 13** Proposed stacked GBDT architecture

## 7 | Exoplanet Identification (Phase II)

After the detection of exoplanets, there might be various instances which are classified as false positives due to non-transiting events such as instrumental noise. Hence, the TCEs need to be vetted to identify Kepler Object of Interest (KOI). This manual process performed by the TCE review team can be automated using machine learning models.

The machine learning analysis is performed on the KOI table that provides the most reliable dispositions and stellar and planetary information such as KOI score, KOI impact, etc. One way to look at the KOI table can be to consider it as a classification problem where the given transiting planet can be classified into one of the classes - false positive or candidate exoplanet. The second way can be to consider this table as a regression analysis problem where we treat KOI score as a target variable whose value ranges between 0 and 1 and on the basis of the predicted value of KOI score we identify whether the given KOI is an actual KOI or a false positive. If the predicted value is close to 1, it designates the KOI as a candidate exoplanet whereas if the value is close to 0, it shows high confidence in the disposition of a KOI as a false positive. For the classification task, we trained - linear model (logistic regression), tree-based models (Decision Trees and Random Forest) and SVM which are explained in the earlier section (Section 6.4). The regression analysis is performed using OLS and Lasso Regression. The explanation of these regression models are given below:

**Ordinary Least Squares Model (OLS)**

In this statistical regression model, the weights $w = (w_0, w_1, w_2, \dots w_n)$ for n training examples are adjusted in such a way that the difference between the sum of squares of observed targets which are there in the dataset and the targets predicted by the linear approximation is minimum. Mathematically, the regression equation can be written in the form of:

$$\min_{n} ||X_n - y||_2^2 \tag{22}$$

**Lasso Regression Model**

The main objective of the Least Absolute Shrinkage and Selection Operator, also known as LASSO is the feature selection and regularization of regression models. It estimates sparse coefficients and gravitates towards solutions having a non-zero value of coefficient and thus reducing the number of features upon which the solution is dependent. Mathematically, the objective function that needs to be minimized is given as:

$$\min_{w} \frac{1}{2n_{samples}} ||X_\omega - y||_2^2 + \alpha ||\omega||_1 \tag{23}$$

## 8 | Habitability Assesment (Phase III)

The habitability assessment on a given exoplanet is performed in the sequence of processes illustrated in the flow chart (Fig. 14). Broadly, the main processes involved are - Preprocessing, Feature Selection, Feature Extraction (computing Absolute Thermal Adequacy (ATA) Score), Bias Handling, and Machine Learning Analysis using Classification and Regression. The preprocessing steps are described in Section 5(b) and Bias handling is done in the same way as in the case of transit signal

detection (Section 6.3). The rest of the processes are elaborated as follows:

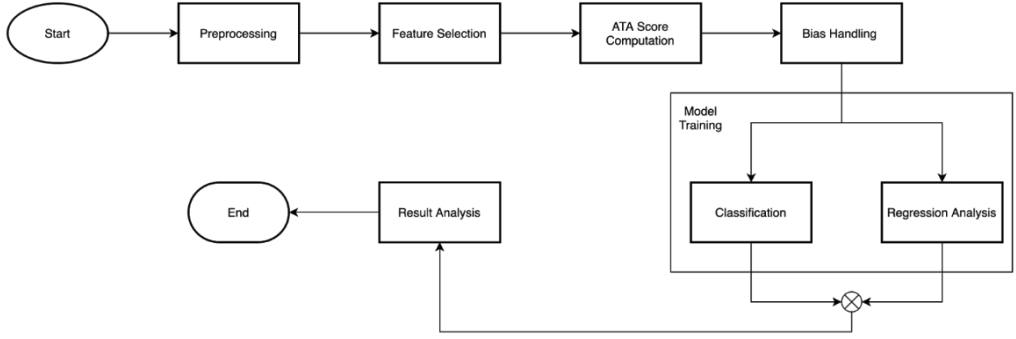

**Fig**. **14** Flow chart of habitability assesment

1. **Feature Selection of Habitability Attributes**

The PHL Catalog data has both categorical as well as continuous features. We adopted a hybrid approach for selecting relevant features from it. Feature Ranking was initially done on the basis of mutual information score (described below) with respect to the target variable. A subset was then prepared by selecting the ten topmost K features from the ranked set of features. Two additional subsets of features were also formed by evaluating feature importance scores using Gradient Boosting Decision Trees and Random Forest models and then assessing the top ten important features from each of them. Finally, the most frequently occurring features among these three subsets were picked out for the classification phase. The complete approach is described in the Algorithm 1.

The Mutual Information is a measure of the similarity between two features and relies on nonparametric methods based on estimating entropy from k-nearest neighbour distances. The Mutual Information between feature U and feature V is given as:

$$MI(U,V) = \sum_{i=1}^{|U|} \sum_{j=1}^{|V|} \frac{|U_i \cap V_j|}{N} \log \frac{N|U_i \cap V_j|}{|U_i||V_j|} \tag{24}$$

where $|U_i|$ is the number of the samples in cluster $U_i$ and $|V_j|$ is the number of the samples in cluster $V_j$.

---

**Algorithm 1.** PHL Feature Selection Algorithm

---

1. **procedure** PHLfeatureSelection(select_K_MI,select_K_GBDT,select_K_RF,count_MI):
   //select_K_MI, select_K_GBDT, and select_K_RF are the vectors of top K selected features based on Mutual Information Score, GBDT feature importance score and Random Forest feature importance score respectively. count_MI is the vector of features' frequency corresponding to select_K_MI feature vector which is initialized as 0.
2.     **for** $feature_i$ **in** select_K_MI **do**
3.       **for** $feature_j$ **in** select_K_GBDT **do**
4.         **if** f$eature_i = feature_j$ **then**
5.           count_MI[index] -> count_MI[index]+1
6.           **for** $feature_k$ **in** select_K_RF **do**
7.             **if** $feature_j = feature_k$ **then**
8.               count_MI[index] -> count_MI[index]+1
9.             **end if**
10.           **end for**
11.         **end if**
12.       **end for**
13.     **end for**
14.     **for** *count* **in** count_MI **do**
15.       **if** *count* > 1 **then**
16.         finalFeatures[index] -> count_MI[index]
17.       **end if**
18.     **end for**
19.     return **finalFeatures**
20. **end procedure**

## 2. Absolute Thermal Adequacy (ATA) Score

A new dimension is introduced in the PHL dataset. We termed it as Absolute Thermal Adequacy Score or ATA score. The reason behind introducing this new feature is to provide a concept of similarity to Earth with regard to equilibrium temperature and distance from the nearest star along with habitability classification of an exoplanet. We intend to find a boundary between potentially habitable and non-habitable exoplanets that separates these classes on the basis of calculated ATA scores. The only features that can be used to calculate this ATA score are the equilibrium temperature and distance from the star as these two factors might provide a reasonable linearly separable relationship between habitable and non-habitable instances.

ATA metric is obtained from the absolute value of the difference between the equilibrium temperature of the exoplanet, $T$ (in K), the equilibrium temperature of the Earth, $T_e$ (in K) and the distance from the nearest star $P_e$ (in AU). Thus, the inputs to the models is in the form of vector $x = [T, P_e/T - T_e|]$. The value of $P_e/|T - T_e|$ is 0 AUK for Earth, thus it can be treated as a central point to the scoring mechanism of the classifiers.

Figure 15 suggests that all the habitable planets lie in the range 35 AUK to 40 AUK whereas the share of non-habitable planets for the same range is very less compared to the habitable planets. Moreover, it can be seen from from Table 1 that the average and median of ATA score among non-habitable planets is very high compared to habitable planets. Also its value varies significantly among non-habitable planets than that of habitable planets. These evidence verifies that ATA score can bring vital information for learning models.

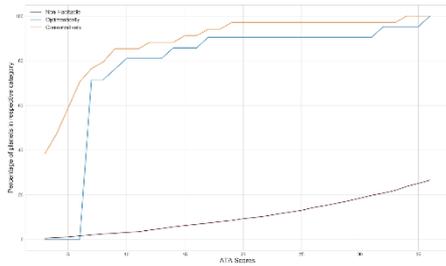

**Fig. 15** Percentage distribution of planets at different ATA scores

**Table 1** Central tendency and variability of ATA score for different habitability

| Habitability type | Mean | Median | Standard Deviation |
|---|---|---|---|
| Non-Habitable | 1.983e+03 | 51.277 | 3.154e+04 |
| Optimistically | 6.457 | 1.823 | 10.17 |
| Conservatively | 6.115 | 4.301 | 6.446 |

## 3. Model Application for Habitability Assessment

Habitability assessment is done for the PHL Catalog data considering two different scenarios. In the first scenario, we treated this assessment as a classification problem where the samples were classified into three different classes - Not Habitable, Conservatively Habitable, Optimistically Habitable. For classification, we have employed the same models as in the case of transit signal detection (Refer Section 6.4). The impact of the ATA metric is also observed in classifying these samples.

In the second scenario, we treated it as a regression problem where we performed regression analysis on Earth Similarity Index (ESI) since this feature is fundamental in determining habitability. The Earth Similarity Index (ESI) (also called easy scale) is an open multiparameter measure of Earth-resemblances for solar or extrasolar planets as a numeric in the range of [0,1] (Schulze-Makuch et al. 2011). The lower bound 0 indicates no similarity to earth whereas the upper bound 1 indicates maximum likeness to Earth.

Mathematically, ESI is a function of two parameters - stellar flux (S) and radius (R), since there is only limited information available about exoplanets. The ESI(S, R) is given by:

$$ESI = 1 - \sqrt{\frac{1}{2}\left[\left(\frac{S - S_\oplus}{S + S_\oplus}\right)^2 + \left(\frac{R - R_\oplus}{R + R_\oplus}\right)^2\right]} \qquad (25)$$

where S is stellar flux, R is the radius, $S_\oplus$ is Earth's solar flux, and $R_\oplus$ is Earth's radius.

However, the goal is to observe the influence of new parameters on ESI. Therefore, we fit a regression curve that can compute the ESI value based on these parameters with minimum error.

From Figure 16, it can be concluded that a high ESI of an exoplanet substantiates with a high possibility of habitability on that planet. Moreover, for superterran planet type, when the planet thermal type is warm, it indicates a high number of optimistically habitable planets and for terran planet type, when the planet thermal type is warm, it indicates a high number of conservatively habitable planets.

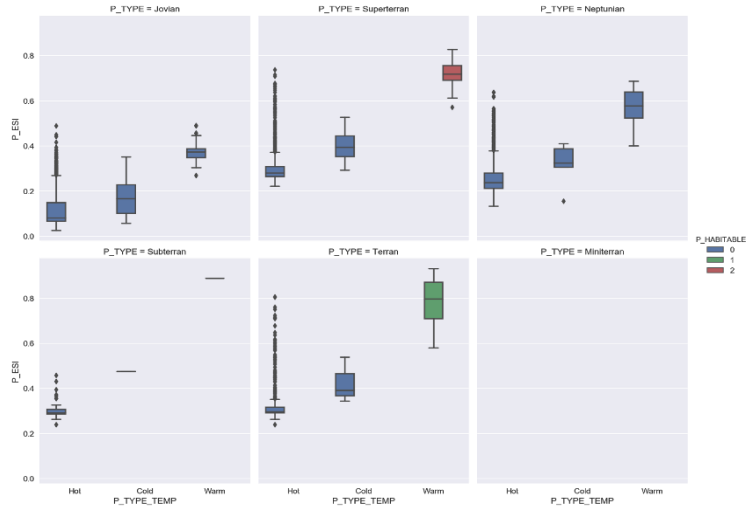

Fig. 16 Earth similarity index vs habitability for different planet and temprature type

## 9 | Evaluation Procedure

The validation of applied models is carried out by 5-fold cross validation scheme. This ensures that each sample of the dataset is put through the training and testing phase at least once. In this way, the unbalanced datasets can be subjected to multiple testing scenarios and thereby avoiding the possibility of biased modelling. Following evaluation metrics are computed in each fold of the cross validation:

$Balanced\ Accuracy = (Recall\ of\ Positive\ Label + Recall\ of\ Negative\ Label)/2$

$Precision = True\ Positive/(True\ Positive + False\ Positive)$

$Recall = True\ Positive/(True\ Positive + False\ Negative)$

$F1\ Score = (2 * Precision * Recall)/(Precision + Recall)$

$False\ Positive\ Rate = False\ Positive/(True\ Negative + False\ Positive)$

We also calculated the ROC AUC score and the PR AUC score for evaluating the ability of our models to distinguish between classes.

## 10 | Result Analysis

Table 2 represents the comparison of models on raw flux for two different dimensionality reduction techniques discussed in this paper by computing Balanced Accuracy, F1 Score, and Positive Predicted Value (PPV). It can be observed that the feature selection outperforms PCA in the case of all the applied models. Moreover, this observation is consistent throughout all bias handling techniques (Refer Table 2 and Fig 17). Therefore, feature selection is considered as the final dimensionality reduction technique on the other extracted signal features for further analysis of models.

**Table 2 :** Comparison between PCA and Feature Selection

|  | | OSS | | | ADASYN | | | Cost Sensitive | | |
|---|---|---|---|---|---|---|---|---|---|---|
|  | Models | Bal. Acc | F1 Score | PPV | Bal. Acc | F1 Score | PPV | Bal. Acc | F1 Score | PPV |
| PCA | SVM | 0.572 | 0.618 | 0.897 | 0.548 | 0.584 | 0.997 | 0.500 | 0.498 | 0.496 |
|  | CART | 0.737 | 0.515 | 0.520 | 0.525 | 0.513 | 0.510 | 0.567 | 0.550 | 0.541 |
|  | RF | 0.659 | 0.533 | 0.548 | 0.500 | 0.498 | 0.496 | 0.512 | 0.520 | 0.663 |
|  | ET | 0.623 | 0.585 | 0.570 | 0.500 | 0.498 | 0.496 | 0.500 | 0.498 | 0.496 |
|  | GBDT | 0.658 | 0.558 | 0.582 | 0.559 | 0.592 | 0.719 | 0.523 | 0.538 | 0.696 |
|  | GBRF | 0.624 | 0.539 | 0.527 | 0.718 | 0.633 | 0.605 | 0.622 | 0.514 | 0.516 |
|  | ANN | 0.500 | 0.498 | 0.496 | 0.622 | 0.622 | 0.622 | 0.500 | 0.498 | 0.496 |
| Feature Selection | SVM | 0.631 | 0.685 | 0.882 | 0.533 | 0.540 | 0.551 | 0.500 | 0.498 | 0.496 |
|  | CART | 0.696 | 0.519 | 0.520 | 0.611 | 0.548 | 0.566 | 0.605 | 0.612 | 0.624 |
|  | RF | 0.756 | 0.645 | 0.604 | 0.570 | 0.598 | 0.701 | 0.523 | 0.538 | 0.621 |
|  | ET | 0.749 | 0.768 | 0.839 | 0.559 | 0.599 | 0.830 | 0.500 | 0.498 | 0.496 |
|  | GBDT | 0..744 | 0.666 | 0.632 | 0..699 | 0.689 | 0.680 | 0.677 | 0.694 | 0.717 |
|  | GBRF | 0.718 | 0.636 | 0.607 | 0.735 | 0.537 | 0.532 | 0.801 | 0.489 | 0.516 |
|  | ANN | 0.681 | 0.616 | 0.586 | 0.742 | 0.628 | 0.589 | 0.500 | 0.498 | 0.496 |

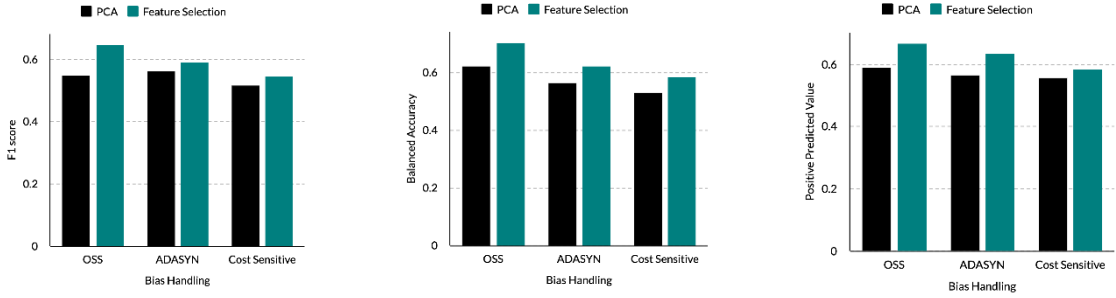

**Fig**. 17 PCA vs Feature selection on the basis of **a:** average F1 score **b:** average balanced accuracy **c:** average positive predicted value

Table 3 depicts the performance of various state-of-art machine learning models along with the proposed Stacked GBDT Model on extracted features - Discrete Fourier Wavelets (DFW) and Power Spectral Density (PSD). In each representation of the signal (DFW and PSD), GBDT is performing better for almost all the evaluation parameters. However, on combining these two representations or characteristics along with time-domain representation, the performance of the proposed stacked GBDT is improved further (Table 3 and Fig 19). Thus, this multi-characteristics/multi-representative model for detecting exoplanets is promising in its performance. Furthermore, it can also be observed that the performance of models including the proposed model on cost-sensitive learning outperformed the models trained on oversampled and undersampled data (Table 3 and Fig 18).

**Table 3 :** Comparison between PCA and Feature Selection

|  |  | OSS | | | ADASYN | | | Cost Sensitive | | |
| --- | --- | --- | --- | --- | --- | --- | --- | --- | --- | --- |
|  | Models | Bal. Acc | F1 Score | PPV | Bal. Acc | F1 Score | PPV | Bal. Acc | F1 Score | PPV |
|  | SVM | 0.844 | 0.805 | 0.789 | 0.500 | 0.598 | 0.496 | 0.631 | 0.698 | 0.942 |
|  | CART | 0.783 | 0.556 | 0.538 | 0.661 | 0.613 | 0.588 | 0.759 | 0.759 | 0.729 |
| Discrete | RF | 0.831 | 0.785 | 0.753 | 0.666 | 0.715 | 0.820 | 0.612 | 0.703 | 0.842 |
| Fourier | ET | 0.830 | 0.804 | 0.789 | 0.714 | 0.783 | 0.922 | 0.667 | 0.743 | 0.964 |
| Wavelets | GBDT | 0.817 | 0.795 | 0.800 | 0.760 | 0.762 | 0.780 | 0.832 | 0.825 | 0.820 |
|  | GBRF | 0.805 | 0.734 | 0.695 | 0.786 | 0.574 | 0.551 | 0.777 | 0.527 | 0.528 |
|  | ANN | 0.500 | 0.498 | 0.496 | 0.714 | 0.783 | 0.922 | 0.500 | 0.498 | 0.496 |
|  | SVM | 0.500 | 0.498 | 0.496 | 0.536 | 0.562 | 0.830 | 0.500 | 0.498 | 0.496 |
|  | CART | 0.566 | 0.488 | 0.506 | 0.501 | 0.500 | 0.502 | 0.506 | 0.506 | 0.507 |
| Power | RF | 0.500 | 0.498 | 0.496 | 0.500 | 0.498 | 0.496 | 0.523 | 0.538 | 0.621 |
| Spectral | ET | 0.525 | 0.543 | 0.696 | 0.500 | 0.498 | 0.496 | 0.500 | 0.498 | 0.496 |
| Density | GBDT | 0.774 | 0.817 | 0.886 | 0..714 | 0.789 | 0.974 | 0.702 | 0.773 | 0.944 |
|  | GBRF | 0.518 | 0.534 | 0.630 | 0.524 | 0.506 | 0.505 | 0.538 | 0.502 | 0.505 |
|  | ANN | 0.500 | 0.498 | 0.496 | 0.500 | 0.498 | 0.496 | 0.500 | 0.498 | 0.496 |
| Proposed Model | Stacked GBDT | 0.844 | 0.859 | 0.886 | 0.820 | 0.824 | 0.828 | 0.809 | 0.853 | 0.916 |

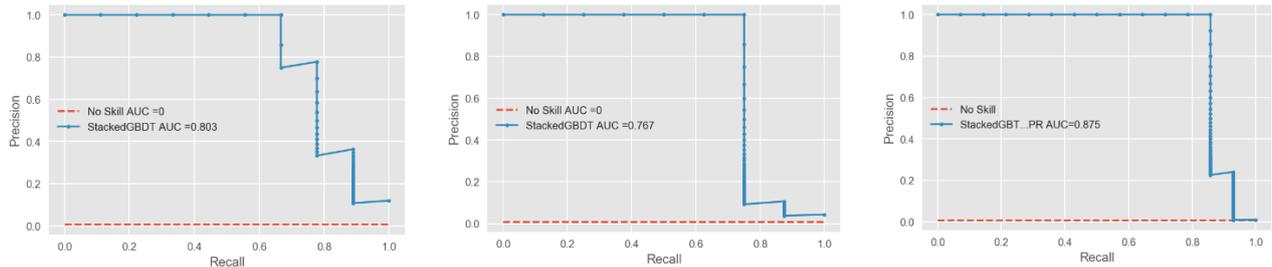

**Fig**. 18 PR AUC of proposed stacked GBDT Model **a:** OSS **b:** ADASYN **c:** Cost sensitive learning

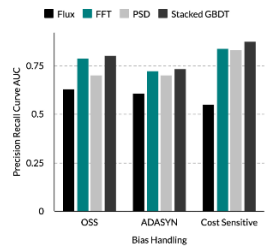

**Fig**. 19 Comparison of maximum PR AUC for each characteristics of light signal with proposed model

Table 4 represents the comparison of various classification models for the exoplanet identification based on F1 score, Accuracy, and PPV. The Random Forest Model is seen to be performing best among other employed models. This can also be confirmed by the PR AUC curve obtained for each of the models (Fig 20). The regression analysis for computing the KOI score is done for two different Models - OLS and LASSO by computing R2 score, MAE, and RMSE (Table 5). It can be observed from the table that OLS is giving more promising results than Lasso regression model.

Table 4 : Classification analysis for exoplanet identification

| Model | Accuracy | F1 Score | PPV |
|---|---|---|---|
| LR | 0.772 | 0.772 | 0.774 |
| NB | 0.626 | 0.578 | 0.746 |
| SVC | 0.623 | 0.576 | 0.709 |
| CART | 0.990 | 0.990 | 0.990 |
| RF | 0.995 | 0.995 | 0.995 |
| ET | 0.882 | 0.883 | 0.894 |

Table 5 : Regression analysis for computing KOI score

| Model | Evaluation Metric | P_ESI |
|---|---|---|
| OLS | R2 Score | 1.000 |
| | MSE | 1.8e-11 |
| | MAE | 1.3e-11 |
| Lasso | R2 Score | 0.807 |
| | MSE | 0.208 |
| | MAE | 0.205 |

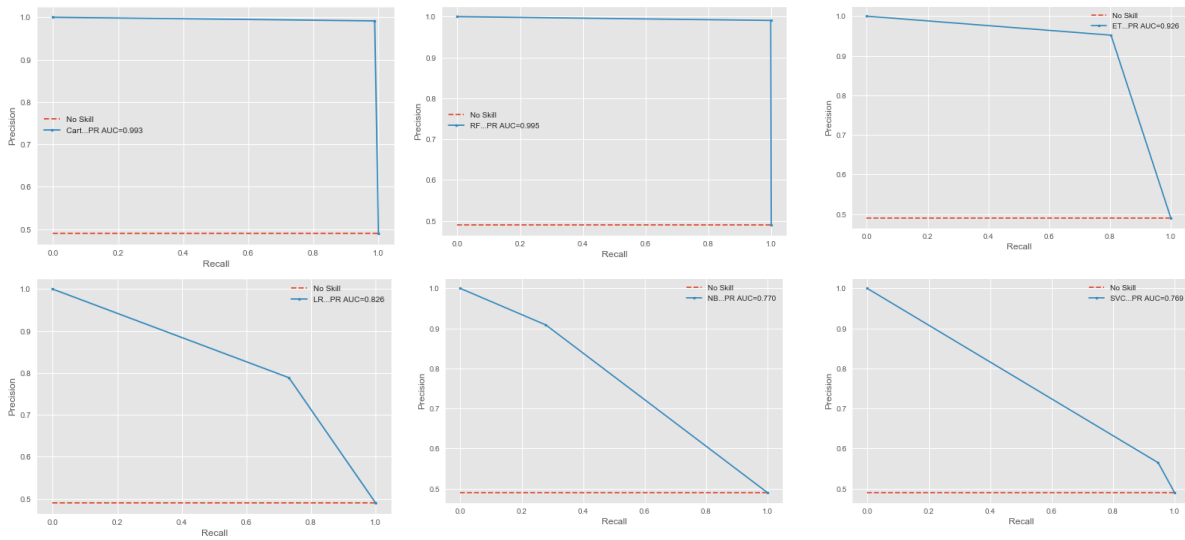

**Fig. 20** PR AUC curve of various models for exoplanet identification

For the Habitability assessment, we tested the performance of employed models for all bias handling techniques considering two scenarios - without ATA Score and with ATA Score. It can be discerned from Table 6 and Fig 21 that including ATA score in the original feature set improves the performance of the models significantly. This corroborates the purpose of devising a feature that can linearly separate between habitable and non-habitable instances. It can be noted that models on undersampled data are performing poorer than cost-sensitive learning and oversampling. However, we cannot claim with confidence which bias handling method is performing better between cost-sensitive and oversampling.

The regression analysis for computing the Earth Similarity Index (ESI) score is done for two different Models - OLS and LASSO by computing R2 score, MAE, and RMSE (Table 7). It can be averred that the regression models are robust. Hence, it confirms our assertion that the ESI score can be obtained with minimum error from the other available features rather than just depending upon stellar flux and planet radius.

Table 6 : Classification analysis for habitability assesment

| | | OSS | | | ADASYN | | | Cost Sensitive | | |
|---|---|---|---|---|---|---|---|---|---|---|
| | Models | Bal. Acc | F1 Score | PPV | Bal. Acc | F1 Score | PPV | Bal. Acc | F1 Score | PPV |
| | LR | 0.430 | 0.410 | 0.423 | 0.444 | 0.457 | 0.480 | 0.333 | 0.331 | 0.329 |
| | SVM | 0.982 | 0.898 | 0.886 | 0.986 | 0.987 | 0.990 | 0.650 | 0.511 | 0.459 |
| Without | CART | 0.929 | 0.565 | 0.499 | 0.958 | 0.967 | 0.980 | 0.946 | 0.958 | 0.980 |
| ATA | RF | 0.938 | 0.735 | 0.676 | 0.962 | 0.975 | 0.991 | 0.949 | 0.962 | 0.980 |
| Score | ET | 0.920 | 0.538 | 0.476 | 0.880 | 0.875 | 0.906 | 1.000 | 1.000 | 1.000 |
| | GBDT | 0.924 | 0.653 | 0.558 | 0.949 | 0.959 | 0.972 | 0.925 | 0.946 | 0.978 |
| | GBRF | 0.932 | 0.587 | 0.496 | 0.950 | 0.978 | 0.989 | 0.895 | 0.911 | 0.948 |

|  | Model | | | | | | | | | |
|---|---|---|---|---|---|---|---|---|---|---|
|  | LR | 0.525 | 0.531 | 0.519 | 0.512 | 0.563 | 0.527 | 0.415 | 0.479 | 0.496 |
|  | SVM | 0.987 | 0.993 | 0.999 | 0.987 | 0.987 | 0.990 | 0.987 | 0.993 | 0.999 |
| With | CART | 0.961 | 0.664 | 0.588 | 0.942 | 0.939 | 0.941 | 0.945 | 0.958 | 0.980 |
| ATA | RF | 0.951 | 0.798 | 0.721 | 0.949 | 0.967 | 0.991 | 0.925 | 0.946 | 0.978 |
| Score | ET | 0.923 | 0.569 | 0.523 | 0.949 | 0.949 | 0.926 | 0.969 | 0.952 | 0.941 |
|  | GBDT | 0.931 | 0.712 | 0.632 | 0.956 | 0.972 | 0.989 | 0.917 | 0.946 | 0.986 |
|  | GBRF | 0.961 | 0.677 | 0.594 | 0.948 | 0.965 | 0.987 | 0.882 | 0.923 | 0.986 |

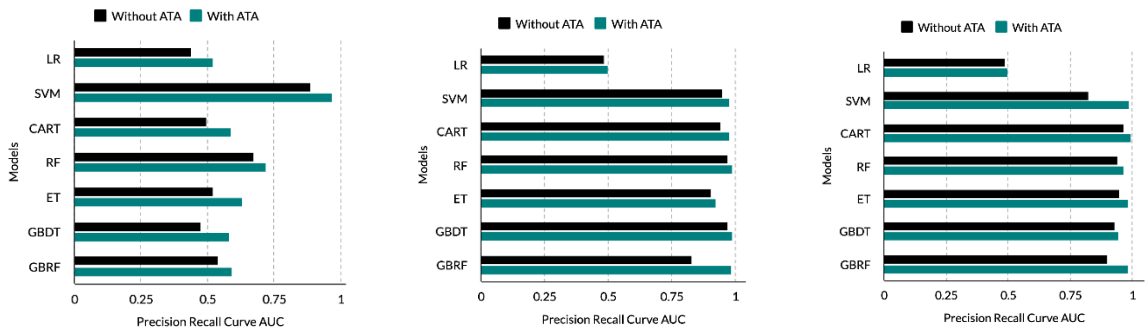

**Fig. 21** Comparison of PR AUC of various models with and without ATA in case of **a:** OSS **b:** ADASYN **c:** Cost sensitive learning

Table 7 : Regression analysis for habitability assesment

| Model | Evaluation Metric | P_ESI | Coefficients |
|---|---|---|---|
| OLS | R2 Score | 0.946 | [-3.309e-01, -6.279e+01, -1.650e+00, |
|  | RMSE | 0.393 | 6.042e+00, -3.201e-01, -2.657e+00, |
|  | MAE | 0.270 | 5.832e-02, 1.114e+00, 3.444e+00] |
| Lasso | R2 Score | 0.770 | [-3.313e-01, -6.276e+01, -1.649e+00, |
|  | RMSE | 0.812 | 6.050e+00, -3.194e-01, -2.658e+00, |
|  | MAE | 0.168 | 5.818e-02, 1.111e+00, 3.445e+00] |

# 11 | Conclusion

Detection, Identification, and Habitability classification of exoplanets are done using manual techniques by astrophysicists and engineers. With an aim to automate these tasks, our study was conducted in three phases in a pipelined manner and various state-of-the-art machine learning and ensemble models were employed by leveraging relevant data in each phase. In the detection phase, we attempt to detect the transiting exoplanet from the dip in the light curve, giving the possibility of the existence of an exoplanetary system around a star. For this, we proposed a multi-representation/multi-characteristic gradient boosting model that considers signals in time-domain, frequency domain, and power spectral domain simultaneously. In the next phase, we automated the triage phase for the identification of exoplanets followed by regression to compute the KOI score which helps in confirming the presence of an exoplanet. In the last phase known as Habitability Assessment, we classified the habitable characteristics of exoplanets into granular clusters of Conservatively habitable, Optimistically habitable, and Non-habitable. We also introduced a new metric called ATA score to classify habitable and non-habitable instances with more robustness. Subsequently, we fitted a regression curve to show Earth Similarity Index (ESI) can be calculated by considering some common attributes rather than relying only on stellar flux and planet radius. The detection and habitability classification was performed after bias handling to tackle severe imbalance.

It is to be noted that detection of transiting exoplanets and habitability determination using machine learning models readily depends on the availability of positive samples (exoplanet and habitable instances). If the positive samples are low in measure, it becomes challenging to train robust models. Since it has been established that finding an exoplanetary system and a habitable planet are scarce phenomena, models should only be trained after a rigorous process of bias handling. In the future, this work can be extended to explore deep learning methods to automate these three phases.

# 12 | References